\pdfoutput=1

\documentclass[acmsmall,screen,nonacm]{acmart}
\usepackage{float}
\usepackage{placeins}

\AtBeginDocument{
  }

\begin{document}

\newcommand{\code}[1]{\nolinkurl{#1}}
\newcommand{\pkg}[1]{\nolinkurl{#1}}
\newcommand{\sem}{\operatorname{Sem}}
\newcommand{\exec}{\operatorname{Exec}}
\newcommand{\Valid}{\operatorname{Valid}}
\newcommand{\Obs}{\operatorname{Obs}}
\newcommand{\Sel}{\operatorname{Sel}}
\newcommand{\ReSA}{\operatorname{ReSA}}
\newcommand{\Gen}{\operatorname{Gen}}
\newcommand{\Cov}{\operatorname{Cov}}
\newcommand{\Dis}{\operatorname{Dis}}
\newcommand{\Pres}{\operatorname{Pres}}
\newcommand{\Atomic}{\operatorname{Atomic}}
\newcommand{\suffix}{\operatorname{suffix}}
\newcommand{\Judge}{\operatorname{Judge}}
\newcommand{\Faithful}{\operatorname{Faithful}}
\newcommand{\Hist}{\operatorname{Hist}}
\setlength{\textfloatsep}{8pt plus 2pt minus 2pt}
\setlength{\floatsep}{8pt plus 2pt minus 2pt}
\title[Approval-to-Execution Repair Analysis]{From Approval to Execution: Reconstruction-Aware Repair Analysis for LLM-Agent Software}
\author{Junchi Zhu}
\author{Zhenguang Liu}
\author{Shaojing Fan}
\author{Jianhai Chen}
\author{Qinming He}
\renewcommand{\shortauthors}{Zhu et al.}

\begin{abstract}
Approval mechanisms have become a primary safeguard for consequential actions in LLM-agent software. Yet the action shown for approval is often not the object ultimately consumed: workflow reload, transcript projection, argument rebinding, and durable-state lookup may reconstruct it before execution. Existing field-flow and check-coverage analyses can establish that expected fields were inspected, but not that the inspected object version reaches the sink or that no replacement intervenes between check and use. Consequently, a locally complete repair may still leave a residual authorization bypass after reconstruction.

We address this problem through three designs. (1) We formulate reconstruction-stable authorization (ReSA) over representation transitions, sink dependencies, consumed versions, and grant scope. (2) We derive obligations that judge candidate repairs and expose residual sink suffixes. (3) We implement the analysis in APAS-Finder and freeze predictions before an independent sink oracle observes execution. Predictions agree with all \code{28} controlled outcomes. Four matched pairs require opposite judgments despite identical field-flow and check coverage; a CodeQL composite recovers all eight when supplied object flow, dominance, interference, and the same grant contract. Two analysts agree on all four model-validation dispositions and \code{19/20} sink dependencies. On released consumers, five repairs prevent \code{60/60} tested out-of-scope effects, while three mechanism contrasts expose the predicted residual effects. Repair sufficiency therefore depends on the reconstructed action consumed, not merely on an approval record or earlier checked representation.
\end{abstract}

\maketitle

\section{Introduction}
LLM-agent software increasingly performs actions with durable consequences: writing files, invoking tools, resuming workflows, and calling external services. Approval is a primary safeguard for these actions. A user reviews a proposal, the system records a grant, and execution resumes under that authority. The safeguard is effective only if the grant remains attached to the action that ultimately reaches the authority-bearing sink.

In practice, approval and execution are rarely separated by one call. The approved action may be persisted and reloaded, projected through a transcript, rebound to tool arguments, or recovered from a durable record. Existing checks typically compare approval metadata or security-relevant fields at one boundary. This direct representation is attractive because it is local, but it hides the cross-representation question: which checked object survives reconstruction and is eventually consumed?

After scrutinizing released approval consumers, we obtained two empirical insights. First, visible approval state can remain intact while a later reconstruction selects a different action: the replacement preserves the expected schema, tool selector, and approval identifier yet changes the affected resource or operation. Second, identical field coverage can lead to opposite outcomes. A check is effective when execution consumes the protected object, but insufficient when a later reread, rebinding, or reselection substitutes another version. Thus, the central question is not only \emph{which fields were checked}, but also \emph{which checked object reaches use and what can intervene along the way}.

The Lobster workflow path exposes the distinction. One execution checks a workflow and later rereads mutable workflow state before shell dispatch; a post-check replacement changes the target that reaches the sink. Another execution consumes the checked immutable snapshot and neutralizes the same replacement. Both expose the same fields and local check, but produce opposite authorization outcomes. They differ in the object that survives to use and in whether reconstruction can intervene between check and use.

\begin{figure}[t]
\centering
\includegraphics[width=.88\linewidth]{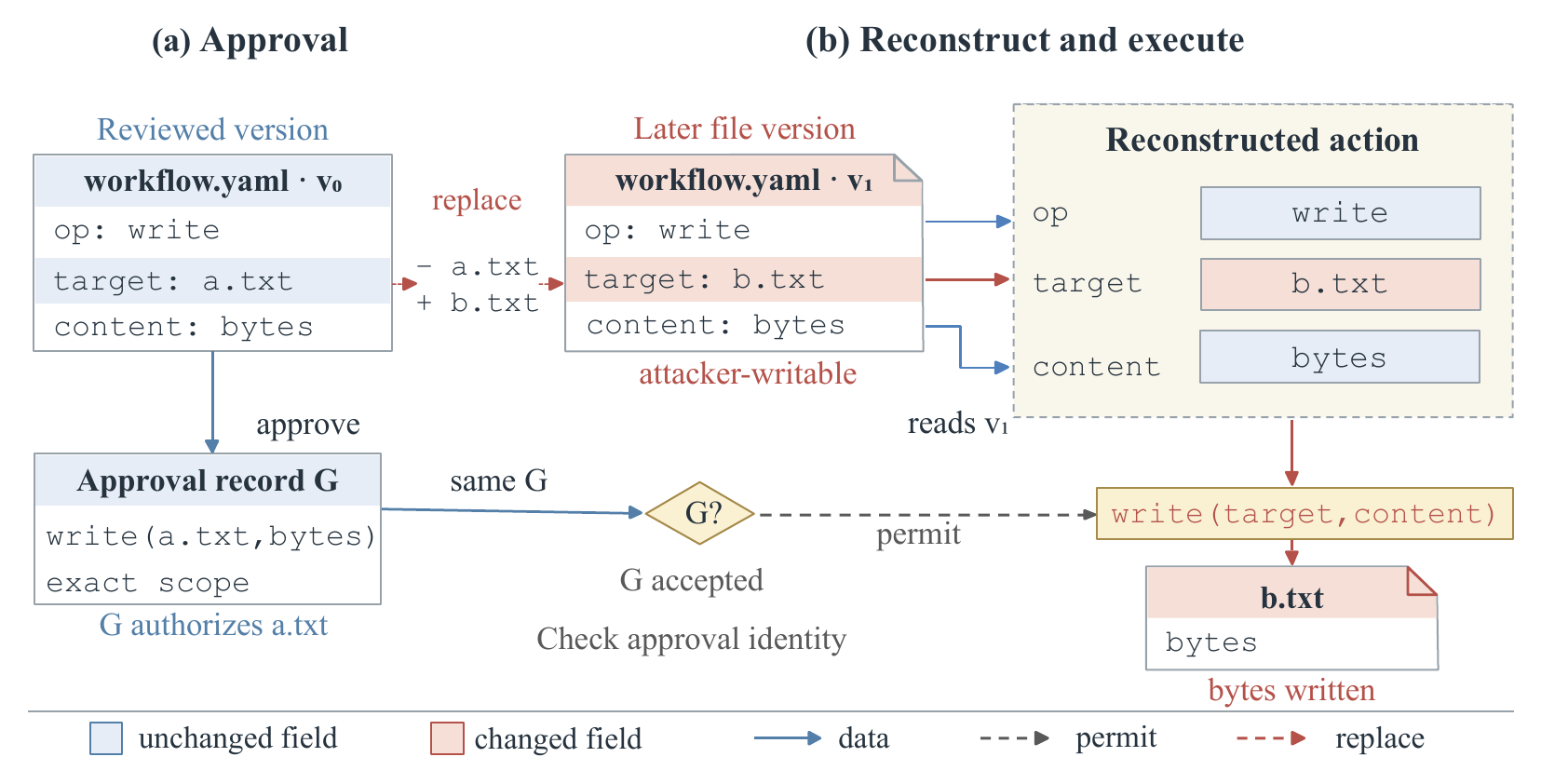}
\caption{Approval-to-execution divergence under post-review reconstruction.
Grant $G$ still authorizes \code{write(a.txt)}, but later mutable state
reconstructs and executes \code{write(b.txt)}.}
\Description{Panel a shows reviewed workflow version v0 and exact approval G
for writing a.txt. Panel b replaces the target in workflow version v1,
reconstructs an action with unchanged operation and content but target b.txt,
retains the accepted approval identifier, and writes bytes to b.txt. A legend
distinguishes unchanged and changed fields, data, permission, and replacement
edges.}
\label{fig:approval-execution}
\end{figure}

Post-check mutation itself is classical, but approval consumers present a distinct analysis target: one semantic action moves across workflow, transcript, tool-call, and durable-state representations while checks reason locally. Consequently, field flow and check coverage cannot determine whether the checked version reaches the sink, whether later selection replaces it, or whether the final action remains in scope. Figure~\ref{fig:approval-execution} illustrates this divergence.

Motivated by these insights, we formulate \emph{reconstruction-stable authorization} (ReSA): every action that reaches the sink under a grant must remain within the grant's authorized semantic scope. We then develop a reconstruction-aware analysis over representation transitions, security-relevant dependencies, trust facts, and existing checks. The analysis generates obligations at selection edges controlled by the attacker and discharges an obligation only when the approved meaning is preserved to the consumed object or atomically validated at use. If a candidate repair fails, the analysis reports the remaining reconstruction suffix to the sink. In this way, repair reasoning shifts from the presence of a local check to the custody of the action that is actually consumed. The formulation builds on interprocedural, object-sensitive, and static security analysis while making consumed versions, authorization scope, and check/use preservation explicit~\cite{repsifds,milanovaobjectsensitivity,livshitsstaticsecurity}.

We implement the analysis in APAS-Finder. Its hybrid front end proposes source-anchored transitions and field facts; a reviewed model supplies the semantic facts that source analysis cannot establish; and the checker automatically derives obligations, repair sufficiency, and residual suffixes. We call a schema-valid replacement that preserves visible approval state but changes the executed meaning an \emph{approval-preserving action substitution} (APAS). APAS-Finder freezes its judgment before an independent executor observes the final sink effect, separating structural prediction from behavioral validation.

We evaluate APAS-Finder on durable grants, approval-tool rebinding, transcript projection, and workflow reload. Predictions match all \code{28} topology--repair outcomes. Four matched pairs hold field flow and check coverage constant while changing consumed object, epoch, preservation, or scope; our analysis distinguishes all four, as does a CodeQL composite when the relevant APIs and grant contract are explicit. Two analysts agree on all four model-validation dispositions and \code{19/20} sink dependencies. Released-code experiments add three blind consumers, five repairs that prevent \code{60/60} tested effects, and three residual-effect contrasts: PromptSpeak partial binding, Lobster check-then-reread, and SSH check-then-reread.

\paragraph{Contributions.}
\begin{itemize}
\item We formulate ReSA and a reconstruction-aware repair analysis that derives obligations relative to approval and judges candidate repairs using consumed object versions, trusted preservation, check/use atomicity, and grant scope.
\item We implement APAS-Finder as a hybrid pipeline for source-anchored candidate extraction, fail-closed model validation, and automatic repair analysis, including residual suffixes for incomplete repairs.
\item We evaluate the analysis through \code{28} controlled repair configurations, four matched abstraction pairs, independent human modeling, three blind released consumers, and five executed repair controls; the controls prevent \code{60/60} tested out-of-scope effects while preserving approved execution.
\end{itemize}

\section{Problem and Threat Model}
\subsection{Security property}
Let $A_r$ denote the action shown during review, $G$ the resulting approval grant, $c$ the execution context, and $A_f$ the final action presented to the authority-bearing sink. The function $\sem(A)$ maps an action $A$ to its normalized security meaning, including the affected resource, operation, principal, and policy context. It deliberately excludes nonce state, expiry, and execution multiplicity, which belong to grant history. Let $S_G$ be the set of security meanings authorized by $G$, and let $\exec(G,c,A)$ be true exactly when the sink executes $A$ under grant $G$ in context $c$. Define the realized semantic set $\mathcal{X}(G,c)=\{\sem(A)\mid\exec(G,c,A)\}$. We require
\[
  \ReSA(G,c) \;\triangleq\; \mathcal{X}(G,c)\subseteq S_G.
  \tag{1}
\]
Broadly, a grant is either exact, with $S_G=\{\sem(A_r)\}$, or scoped over several explicitly bounded meanings. We call Equation~(1) \emph{reconstruction-stable authorization} (ReSA). Approval presence, identifier equality, and schema validity are insufficient because none identifies the action consumed at the sink.

ReSA captures action integrity for one execution rather than execution count. Let $\Hist(c)=\langle e_1,\ldots,e_m\rangle$ be the ordered grant history through the candidate use, $N_G(\Hist)$ its successful uses of $G$, and $b_G\in\mathbb{N}\cup\{\infty\}$ the use budget (one-shot means $b_G=1$). Freshness additionally requires a valid nonce/expiry and $N_G(\Hist(c))\leq b_G$. This separation matters: duplicate execution can violate freshness while preserving the exact approved meaning. Replay is APAS only if it also changes that meaning.

\subsection{Reconstruction-mediated APAS}
\label{sec:apas-definition} A \emph{reconstruction boundary} transforms one action representation into the next using context available after review. Let $A_0=A_r$, and for boundary $i\in\{1,\ldots,k\}$ let $R_i$ be its transformation and $\sigma_i$ its post-review context:
\[
  A_i=R_i(A_{i-1},\sigma_i), \qquad A_f=A_k.
\]
The function $\Obs(G,A)$ returns the approval observables carried with action $A$, such as status, grant identifier, and tool selector. The predicate $\Valid(G,c)$ states that grant $G$ remains valid in context $c$.

\begin{figure}[t]
\centering
\includegraphics[width=.88\linewidth]{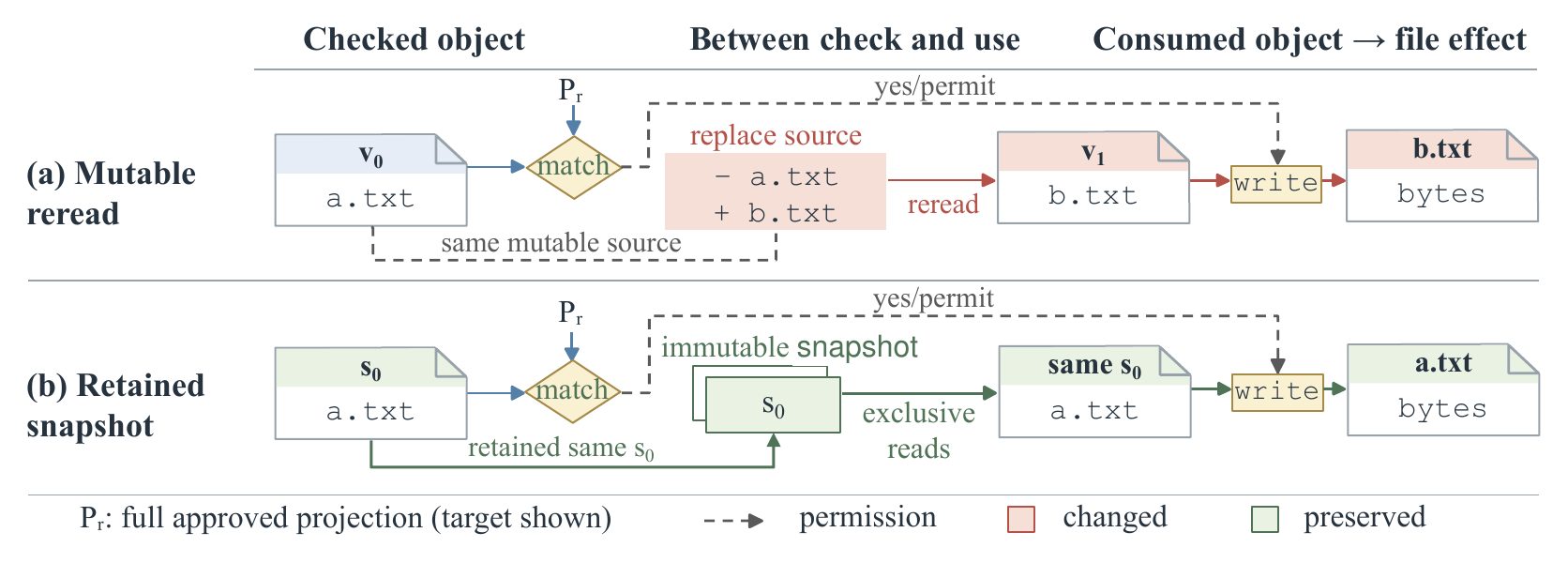}
\caption{Representation transitions and object custody determine whether
approval survives reconstruction. A mutable reread consumes $v_1$ after
checking $v_0$; a retained immutable snapshot consumes the checked $s_0$.}
\Description{Panel a checks target a.txt in object version v0, permits the
operation, then rereads a replaced source as v1 and writes b.txt. Panel b
checks snapshot s0 and requires all later reads to consume that same immutable
snapshot, preserving the write to a.txt. The legend distinguishes permission,
changed values, and preserved values.}
\label{fig:object-custody}
\end{figure}

A schema-valid $A_f$ is a reconstruction-mediated \emph{approval-preserving action substitution} (APAS) when
\[
\begin{gathered}
\Valid(G,c)=\mathsf{true}, \qquad \Obs(G,A_f)=\Obs(G,A_r),\\[-1mm]
\exec(G,c,A_f)=\mathsf{true}, \qquad \sem(A_f)\notin S_G,
\end{gathered}
\tag{2}
\]
and at least one $R_i$ uses attacker-controlled $\sigma_i$ to select a sink-relevant value. Together, these conditions isolate a changed but apparently approved action that reaches the authoritative sink. A semantically unchanged replay, an in-scope change, or a forged packet rejected downstream therefore is not APAS. Figure~\ref{fig:object-custody} contrasts the two decisive cases: rereading mutable state and consuming the protected object.

\subsection{Adversary and trusted base}
The attacker controls one declared post-review input to a consumer: approval arguments, transcript fields, proposal state, or workflow bytes reachable through a shared process or filesystem root. At that boundary, the attacker may submit schema-valid replacements and may retry, reorder, or replay protocol operations. Every evaluated witness identifies the controlled boundary and its deployment precondition. The Lobster witness, for example, requires a local or concurrent workflow writer between halt and resume; this condition does not apply to the other topologies.

The approval store, server-side MAC key, and final executor form the trusted base. The attacker cannot modify them, extract the key, or forge a valid MAC. Cross-process tests additionally assume access to the declared shared root and a known synthetic workflow identifier. Experiments use only synthetic identities, targets, credentials, and sinks.

\subsection{Success oracle and impact}
Operationally, \code{mutationReached} means that the sink executed an action $A_f$ with $\sem(A_f)\notin S_G$. Changed target, content, operation, caller, or policy fields can redirect a write or another privileged effect. Lifecycle and replay mutations are scored separately against freshness rather than folded into $\sem(A)$. The concrete impact depends on the downstream tool. Our experiments establish reachability only to synthetic local sinks; an approval response or a mutation rejected before the sink does not satisfy the oracle.

\section{Methodology}
\label{sec:method}
\subsection{Reconstruction-aware analysis}
\paragraph{Overview.}
Fundamentally, repair analysis must determine whether every attacker-controlled selection of a sink-relevant value is protected before use. Rather than treating each checked representation independently, APAS-Finder follows the action to the object consumed at the sink. It first reconstructs the representation path, then generates and discharges approval-relative obligations through trusted preservation or atomic validation, and finally reports repair sufficiency and every surviving suffix. The abstraction therefore models object custody instead of check presence alone.

\paragraph{Graph and inputs.}
We represent a consumer as a directed representation-transition graph $G_R=(V,E)$. Each vertex in $V$ is an action representation, and each edge $e\in E$ transforms its source representation into its destination. A path starts at the reviewed action $A_r$ and ends at the final sink action $A_f$. Let $E_m\subseteq E$ contain edges that read mutable post-review context, and let $E_a\subseteq E_m$ contain those whose context the stated attacker controls.

For every edge $e$, let $\Sel(e)$ be the set of fields whose values the edge can select from its inputs. The analyst supplies $\Sel(e)$, the edge's write set, mutability, attacker control, and normalization. The analyst also supplies the set $D$ of fields on which the sink's security meaning depends, the grant scope $S_G$, and source or trace anchors for these facts. Expected classifications, repair locations, residual paths, and sink outcomes are not inputs. This formulation uses graph-reachability dataflow analysis but specializes its facts to approval-relative reconstruction~\cite{repsifds,fisleraccesscontrol}.

The supplied facts describe local implementation events rather than repair outcomes. In contrast, the checker composes them across joins, reselections, and repair placements. This has two consequences: it determines whether a repair closes every sink path and locates the first uncovered suffix when it does not. Thus, the seven RQ1 repairs share graph facts yet produce different judgments.

Operationally, annotators obtain $\Sel(e)$ from assignments, argument construction, deserialization, and lookups, and obtain $D$ by tracing the sink's authority-bearing arguments and scope predicates backward. Every fact names a source location or observed read; missing or conflicting evidence is unknown. AST or dataflow tools can propose reads, but analysts confirm scope, attacker control, trusted preservation, and check/use atomicity. Atomicity is accepted only when validation and use are linearized by the same transaction or lock, or occur in one interference-free critical section over the consumed object. Consumption of an authenticated immutable object without fallback rereads is recorded separately as trusted preservation. Otherwise the corresponding fact is \emph{unknown}.

The complete security projection $P_D(A)$ is the normalized tuple of fields in $D$, containing exactly the information that decides $\sem(A)\in S_G$. Its normalization rules and version are inputs and must agree at approval and use; the analyzer does not infer semantic equivalence from byte equality.

\paragraph{Hybrid front end and responsibilities.}
APAS-Finder separates modeling and inference into three stages, each addressing a different source of uncertainty. \code{extract} proposes anchored transitions, field accesses, sink calls, and immutable copies. \code{validate} rejects contradictions and queues unresolved scope, attacker control, preservation, atomicity, and opaque links for review. Given a valid model, \code{analyze} generates obligations, evaluates repairs, and emits anchored residual suffixes. Unknown facts remain unknown; a reviewed version-pinned model can then be rechecked as code or repairs change.

\paragraph{Authorization obligations.}
For an edge $e$, define the generated obligation set
\[
  \Gen(e)=\{(e,f)\mid e\in E_a\land f\in\Sel(e)\cap D\}.
  \tag{3}
\]
Intuitively, each $o=(e,f)\in\Gen(e)$ records that edge $e$ can select sink-relevant field $f$ from attacker-controlled context. The value must be protected before use; execution determines whether the potential gap is feasible.

\paragraph{Repair discharge.}
Let $H$ be the protections provided by a candidate repair. For protection $h\in H$ and path $\pi$, $\Pres(h,\pi)$ means that the sink consumes the checked immutable version or an authenticated immutable copy preserving every field in $D$. Content equality alone does not establish preservation.

A protection covers field $f$, written $f\in\Cov(h)$, when it checks the authorization-relevant value of $f$. The predicate $\Atomic(h,\pi)$ means that $h$ validates the complete projection of the version consumed on $\pi$ and uses it in one interference-free epoch. Let $H_\pi[e,\mathrm{sink}]$ contain protections after edge $e$ and no later than sink use. We define
\[
\begin{aligned}
\Dis_H^\pi(e,f)\ \Longleftrightarrow\ {}
&\exists h\in H_\pi[e,\mathrm{sink}]\ \text{such that}\\[-1mm]
&f\in\Cov(h)\ \land\
\bigl(\Pres(h,\pi)\ \lor\ \Atomic(h,\pi)\bigr).
\end{aligned}
\tag{4}
\]
Equation~(4) provides two complementary mechanisms: preserve a trusted checked version to use, or atomically validate and consume the complete final projection. The first controls object identity; the second excludes intervening writes, rebinding, rereads, and reselection. Hence identical field coverage can yield opposite judgments when only one repair protects the object that reaches use~\cite{milanovaobjectsensitivity,schwartztaint}.

\paragraph{Analysis procedure and outputs.}
Let $\Pi$ be the declared sink paths and $Q\subseteq D$ the approval projection. For path $\pi=(e_1,\ldots,e_k)$, the analyzer propagates outstanding obligations and returns their sink suffixes:
\[
\begin{aligned}
U_0^\pi&=\varnothing,\\
U_i^\pi&=\{(e,f)\in U_{i-1}^\pi\cup\Gen(e_i)\mid
\neg\Dis_H^\pi(e,f)\},\\
\mathcal{R}(H)&\triangleq\bigcup_{\pi\in\Pi}
    \{(o,\suffix(\pi,e_o))\mid o\in U_k^\pi\}.
\end{aligned}
\tag{5}
\]
Equation~(5) answers both whether an obligation survives and where its bypass begins: union retains uncovered predecessors at joins, and $\mathcal{R}$ pairs each survivor with its generating-edge suffix. The final judgment is
\[
\Judge(H)=
\begin{cases}
\mathsf{unknown}, & D\nsubseteq Q\ \lor\ \text{required facts are unknown},\\
\mathsf{sufficient}, & \mathcal{R}(H)=\varnothing,\\
\mathsf{insufficient}(\mathcal{R}(H)), & \text{otherwise}.
\end{cases}
\tag{6}
\]
Equation~(6) distinguishes three outcomes: unresolved facts yield \textsf{unknown}, an empty residual set yields \textsf{sufficient}, and every survivor is returned with its suffix. The result is relative to declared paths, trust, and atomicity; execution at the final sink determines whether a candidate residual path is feasible.

\paragraph{Worked derivation.}
Consider \code{review->resume->reload->sink}. Edge $e_1$ reloads the workflow, after which $e_2$ resolves target, content, and operation from a mutable reference. Equation~(3) generates four obligations. Freezing at $e_1$ discharges $(e_1,\mathtt{workflow})$, but recurrence~(5) carries the three obligations generated at $e_2$ to the sink, so Equation~(6) judges the repair insufficient. An atomic final check discharges all four, as does \code{freeze_all_inputs} when execution consumes only checked immutable inputs. Thus, protecting the first reconstructed object need not protect values selected later. Figure~\ref{fig:obligation-derivation} shows the derivation.

\paragraph{Conditional guarantee.}
An instance $\Gamma$ is \emph{faithful} when $\Pi$ covers every feasible sink path, $D$ decides membership in $S_G$, $\Sel$ and $E_a$ cover every controlled selection in $D$, and all preservation and atomicity facts are correct. Then
\[
  \Faithful(\Gamma)\land\Judge_\Gamma(H)=\mathsf{sufficient}
  \;\Longrightarrow\; \ReSA_\Gamma(G,c),
  \tag{7}
\]
where $\Gamma=(G_R,D,Q,\Pi,E_a,\Sel,S_G,\Pres,\Atomic)$. Under these assumptions, every out-of-scope action differs on $D$ and leaves an obligation unless trusted preservation or atomic validation discharges it. Thus, sufficiency over a faithful reviewed model implies ReSA for that model.

\FloatBarrier
\subsection{Prediction protocol}
\label{sec:prediction-protocol}
\begin{figure}[t]
\centering
\includegraphics[width=.90\linewidth]{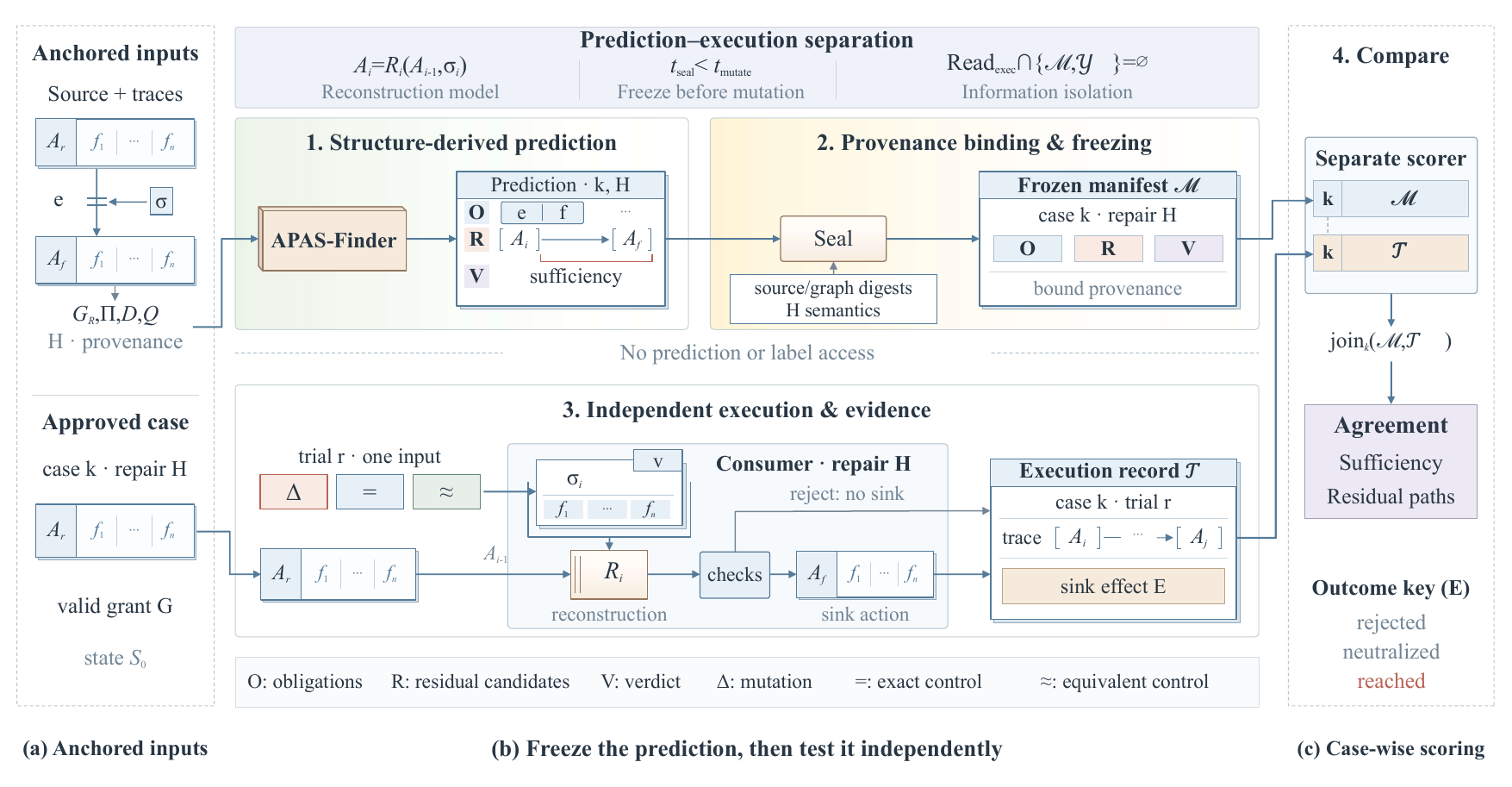}
\caption{Structure-derived prediction and independent execution validation.
The executor does not read frozen predictions or labels; a separate scorer
compares them with final sink effects.}
\Description{Panel a supplies anchored representations and a reviewed action.
Panel b separates a structure-based predictor and sealed manifest from an
executor that receives only the approved case and mutation. Panel c joins the
manifest and execution record by case identifier; the executor reads neither
the manifest nor expected labels, and the scorer compares the sink effect with
the frozen prediction.}
\label{fig:prediction-validation}
\end{figure}

\paragraph{Prediction--execution separation.}
To separate prediction from observed behavior, the predictor writes a sealed manifest $\mathcal{M}$ containing graph digests, intervention semantics, obligations, sufficiency, and residual candidates; the scorer separately retains expected labels $\mathcal{Y}$. A stateful executor constructs approved state, applies a fresh mutation, runs reconstruction and checks, and records the sink effect without reading either file. A third program joins prediction and observation by case identifier. Exact and semantically equivalent controls traverse the same executor. Figure~\ref{fig:prediction-validation} shows these boundaries.

\begin{figure}[t]
\centering
\includegraphics[width=.88\linewidth]{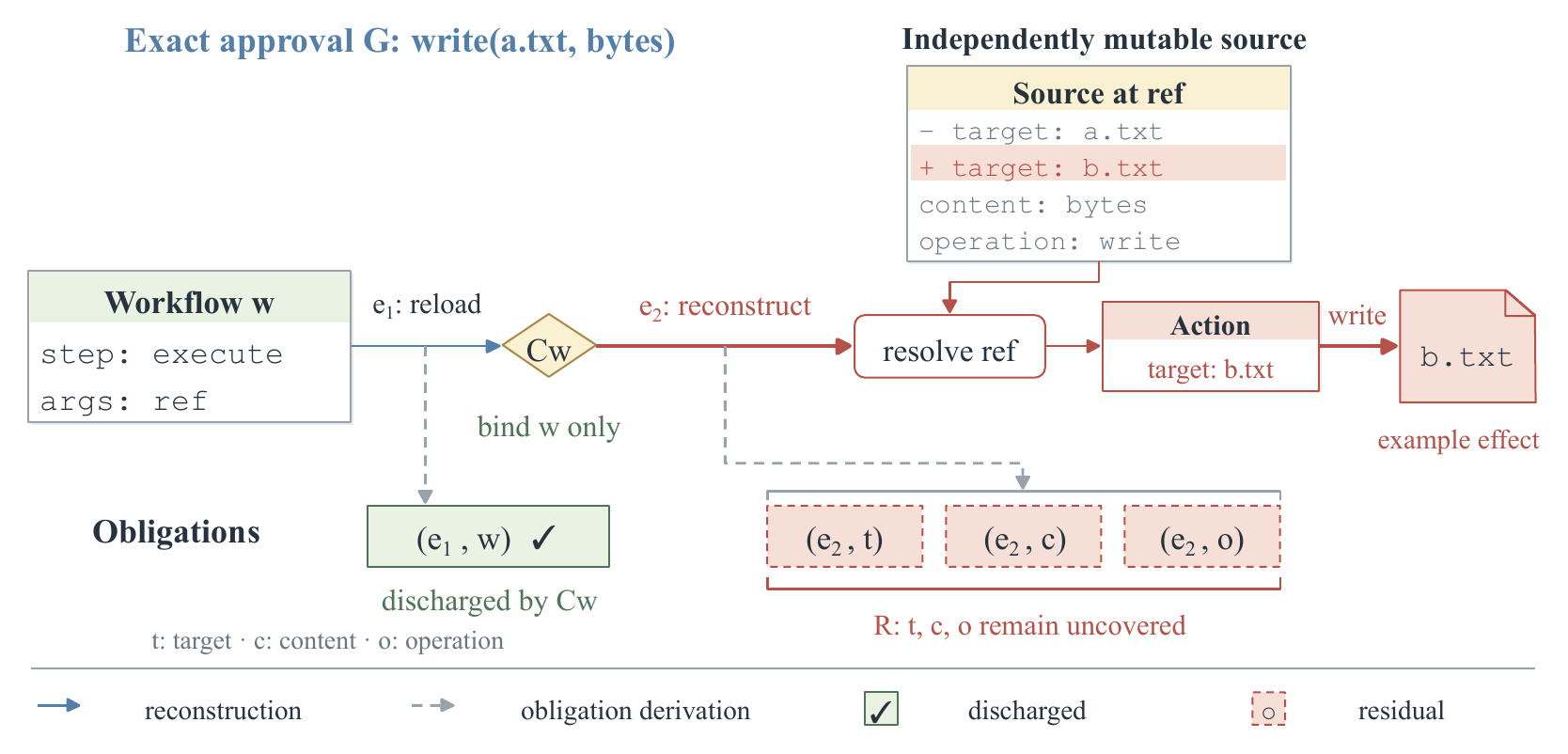}
\caption{Reconstruction path and authorization-obligation derivation. Binding
the workflow at $e_1$ discharges $(e_1,w)$, but resolving an independently
mutable source at $e_2$ leaves three residual obligations.}
\Description{An exact grant authorizes writing a.txt. Workflow w reloads a
reference at edge e1, where workflow obligation w is discharged. Edge e2
resolves an independently mutable source that changes the target to b.txt,
constructs the final action, and writes b.txt. Residual obligation boxes mark
target, content, and operation as uncovered.}
\label{fig:obligation-derivation}
\end{figure}

Figures~\ref{fig:prediction-validation} and \ref{fig:obligation-derivation} make the information boundary explicit. The analyst supplies anchored implementation facts; the analyzer derives and freezes obligations, residual candidates, and repair verdicts before replacement.

\subsection{Study instantiation}
\paragraph{Semantic controls and fail-closed boundary.}
The semantic contract includes exact normalized execution, explicitly scoped reuse, incomparable cross-target substitution, and a mutable edge outside attacker control. These controls separate authorization from raw field equality and attacker reachability from mere mutability. Placement controls exercise an early check followed by reconstruction, a late atomic check, and a late check/use race.

Projection completeness is judged relative to declared sink dependencies. If a graph-declared sink-influencing field is absent from the approval projection, the analysis returns \code{unknown} for an incomplete projection rather than certifying the repair. The check therefore covers graph-visible dependencies; graph construction supplies the dependency universe. Analysis-time abstention avoids an unsupported safety conclusion, whereas a runtime fail-closed guard prevents execution.

\paragraph{Released paths.}
For each version-pinned consumer, we reconstruct the approval-to-execution path from published code. The audit identifies the approval-bearing representation, post-approval reads or writes selecting security-relevant fields, the first boundary accepting the reconstructed action, and the authority-bearing sink. For each transition, we record source anchors, checked and consumed objects, attacker control, trusted preservation, and check/use atomicity. These facts instantiate the graph without encoding a repair outcome.

We classify sink evidence as package-owned, delegated, recorder-backed, synthetic child process, or packet-only. An approval response alone is not impact: a dynamic witness must carry an out-of-scope action to the declared sink. Static auditing establishes the released path and its assumptions; dynamic execution tests reachability. Evidence classes remain separate in aggregate reporting.

The experiments invoke shipped approval APIs with synthetic review events. This isolates reconstruction behavior from operator-interface effects and does not model whether a human would approve the original action. The artifact records version identifiers, source anchors, graph facts, and sink ownership for every consumer.

\paragraph{Generative-AI assistance.}
ChatGPT and Codex were used during research to suggest experimental contrasts, draft and revise JavaScript runners and scorers, review consistency, draft figures, and edit prose. Authors selected the research questions and accepted or rejected suggestions; inspected source anchors and every retained code change; froze predictions before execution; ran the experiments; and checked raw sink effects, aggregate counts, and citations. No model response serves as a runtime oracle, observation, ground-truth label, or statistical sample. Deterministic checked-in programs regenerate reported aggregates from the frozen records. One analyst's review notes were converted to the submission schema with LLM formatting assistance; retained source anchors and unknown facts, rather than model-generated labels, determine the reported human result.

\section{Evaluation}
To evaluate APAS-Finder, we ask whether reconstruction structure predicts repair outcomes, which authorization facts make those predictions possible, and whether frozen diagnoses transfer to released consumers. We organize the evidence into finite-model agreement, mechanism-isolating controls, and released-code execution, with a separate unit and inference scope for each.

\paragraph{Research questions.}
\begin{itemize}
\item \textbf{RQ1:} Can reconstruction structure predict repair sufficiency and residual bypass paths before execution?
\item \textbf{RQ2:} Which authorization-relevant facts are required, and how do incomplete inputs affect the resulting judgments?
\item \textbf{RQ3:} Do frozen diagnoses and repair predictions transfer to released consumers?
\end{itemize}

We first establish the subjects, prediction protocol, execution oracle, and repair controls. We then answer each research question with the corresponding evidence unit.

\subsection{Experimental Settings}
\paragraph{Subjects and experimental units.}
RQ1 evaluates seven repairs on four frozen graphs: 28 configurations and 140 edge--field executions. RQ2 uses structure-ablated baselines, eight matched cases, obligation and projection omissions, and six graph-fact perturbations. RQ3 transfers predictions to three blind consumers, executes five reference controls, and tests three residual-effect contrasts: PromptSpeak partial binding, Lobster check-then-reread, and SSH check-then-reread. Repetitions test stability, not independent samples.

\paragraph{Independent graph-construction protocol.}
Two analysts independently reviewed APAS-Finder's prepopulated candidates for four version-pinned consumers. They received source, task specifications, and extraction queues, but not author graphs, predictions, outcomes, or each other's submission. They source-anchored every decision, retained unknowns, and froze their models before validation. We report pre-adjudication results and self-reported active time.

\paragraph{Prediction protocol.}
The evaluation follows Section~\ref{sec:prediction-protocol}. Predictions are frozen before execution and joined to observations by case ID. Released subjects span package-owned, simulated, synthetic child-process, delegated, recorder-backed, and packet-only sinks; we report these evidence classes separately. Contract and legacy-runner checks remain artifact regressions rather than evaluation units.

\paragraph{Execution oracle.}
For each run, the adapter creates an approved action $A_r$ and a schema-valid candidate $A'$ that changes one security-relevant component. The candidate traverses the normal approval path, and the adapter observes $E$ at the final boundary. The Boolean $executionOccurred$ records whether the sink produced an effect $E$; $mutationReached$ records whether that effect falls outside the approved scope. The primary failure predicate is
\[
\begin{aligned}
mutationReached ={}& executionOccurred\\
&\land \sem(E)\notin S_G.
\end{aligned}
\]
For exact approvals, $S_G$ is the singleton semantic action reviewed by the adapter; scoped controls may accept changed actions that remain inside $S_G$. The runner separately records mutation neutralization when the original action executes, scoped reuse when a changed action remains inside $S_G$, and rejection before executor when no execution occurs. Each mutation is repeated five times with a fresh synthetic root. The artifact enumerates target, content, operation, scope, identity, lifecycle, and replay families. APAS-Finder generates type-preserving replacements from semantic field names for common target, content, operation, scope, caller, and session cases; adapters may add sink-specific overrides when one logical operation expands to multiple package operations. This keeps mutation generation reusable while making package-specific assumptions explicit, following mutation testing's emphasis on explicit operators and observable outcomes~\cite{jiaharmanmutation}.

The oracle records preserved approval observables, the substituted projection component, the first accepting reconstruction boundary, the final sink effect, and the scope verdict. These fields separate reconstruction-mediated APAS from valid scoped reuse, prompt-induced selection without a changed grant, stale verdicts, duplicate execution, and packet-only tampering.

\paragraph{Repair controls.}
\label{sec:reference-repair} The reference repairs define a compact design space for preserving approval through reconstruction. A final-boundary repair binds a canonical security projection of $A_r$ to a fresh nonce, expiry, principal/session, approval identifier, and tool/policy version; the executor recomputes the projection from $A_f$ immediately before the consequential operation. A snapshot repair instead authenticates the representation that later stages consume and forbids fallback reads or redirection to the original mutable source. A composed repair may use an early projection for efficient rejection and a final atomic check for the fields reconstructed later. The analyzer selects among these candidates by which obligations they discharge, allowing the repair to match the consumer's reconstruction path.

Repairs fail closed on projection, identity, expiry, tool/call, or nonce mismatch. A final check requires complete atomic use; a snapshot requires all later reads to consume the authenticated object. Both require complete projection, canonicalization agreement, trusted custody, and atomic lifecycle state.

\subsection{Repair Prediction and Residual Localization (RQ1)}
RQ1 asks whether declared reconstruction structure predicts both a candidate repair's outcome and the bypass suffix that remains before the repair is executed. We evaluate the complete intervention matrix and then remove position, atomicity, or later-reconstruction facts from the analysis. The resulting 28 configurations form a factorial mechanism test over four authored graphs and seven interventions; they are not 28 independent programs.

\subsubsection{Intervention outcomes}
\begin{figure}[t]
\centering
\includegraphics[width=.88\linewidth]{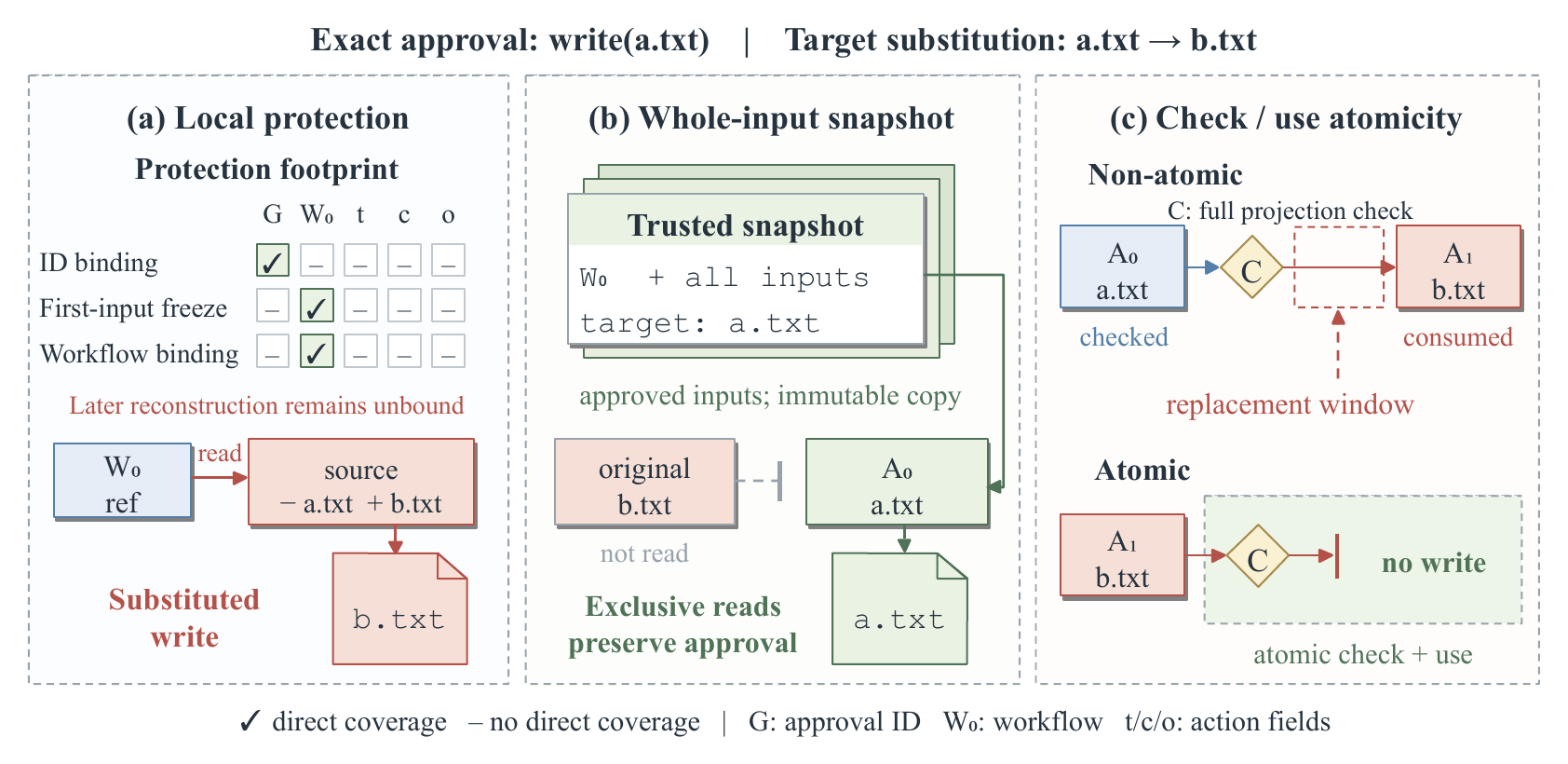}
\caption{Repair mechanisms and topology-dependent sufficiency. Local
protection leaves reconstruction unbound; snapshot consumption and atomic
check/use protect the object reaching the sink.}
\Description{Panel a shows identifier, first-input, and workflow bindings that
do not directly cover later target, content, and operation reconstruction,
allowing a substituted write. Panel b consumes only a trusted snapshot of the
workflow and all inputs, preserving a.txt despite mutation of the original.
Panel c contrasts a non-atomic projection check with an atomic check and use
that prevents the out-of-scope write.}
\label{fig:repair-mechanisms}
\end{figure}

Figure~\ref{fig:repair-mechanisms} summarizes the topology-dependent repair judgments; the case-level evaluation below retains all 28 interventions. The seven intervention IDs are:
\begin{center}
\small
\code{approval_id_only}; \code{first_accepting_projection};
\code{freeze_first_input}; \code{final_projection_atomic};\\
\code{final_projection_racy}; \code{freeze_all_inputs};
\code{first_plus_final}.
\end{center}
A no-binding control is reported outside this denominator. The core evaluation asks whether reconstruction structure predicts intervention outcomes before execution. From the four frozen graphs, the analyzer generated 20 obligations and evaluated seven interventions per topology, yielding 28 topology--intervention cases and 140 edge--field mutation executions. The stateful executor did not read the prediction or label files. Prediction and execution agreed on repair sufficiency in \code{28/28} cases and on the exact residual-path set in \code{28/28}. Exact and key-reordered semantic-equivalent actions remained executable throughout.

Two observations explain the agreement. First, atomic final projection blocked every mutation in \code{4/4} topologies, whereas a non-atomic final check left an executable substitution in \code{4/4}. Early projection and first-input freezing worked for the two single-mutable-edge graphs but failed when transcript or workflow reconstruction selected another mutable value. Lobster's checked snapshot is therefore \code{freeze_all_inputs}, not \code{freeze_first_input}: every later workflow read and resolved action value must come from it. Second, the scorer recovered the exact suffix for every insufficient repair, not only its binary outcome. The results connect each repair mechanism to the reconstruction it does or does not dominate.

We froze the protocol and prediction matrix before execution. The manifest records graph-fact, intervention-semantics, and source digests. The executor writes reconstruction traces and realized sink effects, and a separate scorer joins prediction and observation by case identifier. The executor distinguishes \emph{rejected} (no sink execution), \emph{neutralized} (the approved action executes from protected state), and \emph{reached} (an out-of-scope action executes). Both rejection and neutralization preserve authorization, but only the former has no sink effect. This is independent behavioral evidence produced under the shared declared semantics.

\subsubsection{Comparison with structure-ablated rules}
We compare four analyzers under a common output contract: repair sufficiency, residual path, and repair location. An unavailable output is scored as an abstention. All methods receive the same topology and intervention descriptors; none reads outcomes. \emph{Final-sink-only} accepts only a complete atomic final check; \emph{atomicity-blind final-binding} accepts any complete final projection; \emph{first-edge} accepts protection at the first mutable edge and assumes later preservation. ReSA applies the reconstruction-aware obligation rules defined in Section~\ref{sec:method}. Accuracy counts non-abstaining verdicts; abstention is reported separately.

Table~\ref{tab:repair-baselines} reports the complete confusion counts. All methods make 28 binary predictions; only diagnostic outputs abstain. Atomicity-blind is false-safe on four racy checks. First-edge is false-safe where \code{history->sink} or \code{reload->sink} remains uncovered; the artifact retains every false-safe and false-unsafe case ID.

\begin{table}[t]
\centering
\caption{Repair-sufficiency predictions on the 28 frozen
topology--intervention cases. FS and FU denote false-safe and false-unsafe
judgments. Residual reports exact residual-path output.}
\label{tab:repair-baselines}
\scriptsize
\setlength{\tabcolsep}{3.2pt}
\begin{tabular}{lrrrrrl}
\toprule
Analyzer & TP & TN & FS & FU & Accuracy & Residual \\
\midrule
Final-sink-only & 8 & 12 & 0 & 8 & \code{20/28} & abstain \code{28/28} \\
Atomicity-blind & 8 & 8 & 4 & 8 & \code{16/28} & abstain \code{28/28} \\
First-edge & 8 & 8 & 4 & 8 & \code{16/28} & abstain \code{28/28} \\
Reconstruction-aware & 16 & 12 & 0 & 0 & \code{28/28} & exact \code{28/28} \\
\midrule
\multicolumn{7}{l}{\emph{Reconstruction-aware breakdown by topology}} \\
Durable-grant consumer & 5 & 2 & 0 & 0 & \code{7/7} & exact \code{7/7} \\
Approval-tool rebinding & 5 & 2 & 0 & 0 & \code{7/7} & exact \code{7/7} \\
Transcript projection & 3 & 4 & 0 & 0 & \code{7/7} & exact \code{7/7} \\
Workflow replacement & 3 & 4 & 0 & 0 & \code{7/7} & exact \code{7/7} \\
\bottomrule
\end{tabular}
\end{table}

The topology rows expose the composition behind the aggregate: the first two graphs contain five sufficient and two insufficient interventions, while the two later-reconstruction graphs contain three sufficient and four insufficient interventions. The aggregate \code{28/28} agreement is therefore a complete factorial mechanism result, not an estimate from 28 independent consumers; its two-sided Wilson 95\% interval is approximately $[0.88,1.00]$.

To hold the security effect constant across representations, we repeated the matrix with one canonical mutation of target, content, and operation. All \code{28/28} exact controls remained valid. In contrast, atomic final projection blocked the mutation in \code{4/4} topologies, while the racy check allowed it in \code{4/4}. The outcome therefore follows repair placement and reconstruction rather than topology-specific field names.

\paragraph{Answer to RQ1.}
Reconstruction structure predicts not only whether each of the \code{28} repairs is sufficient, but also the exact suffix through which every insufficient repair remains bypassable.

\subsection{Required Authorization Information (RQ2)}
RQ2 asks which authorization facts change a repair judgment when field flow and check coverage remain fixed. Following an ablation design, we vary one semantic dimension or one declared input at a time and observe the resulting prediction and sink effect.

\subsubsection{Effect of object, version, epoch, and scope facts}
Four matched pairs fix mutable fields, sink dependencies, and check coverage while varying check/use epoch, consumed object, later reselection, or grant scope. The contrast is decisive: a field-flow/check-coverage rule scores \code{4/8} and separates \code{0/4} pairs, whereas ReSA scores \code{8/8} and separates \code{4/4}. Prediction, execution, and scoring remain separate programs.

\begin{figure}[t]
\centering
\includegraphics[width=.88\linewidth]{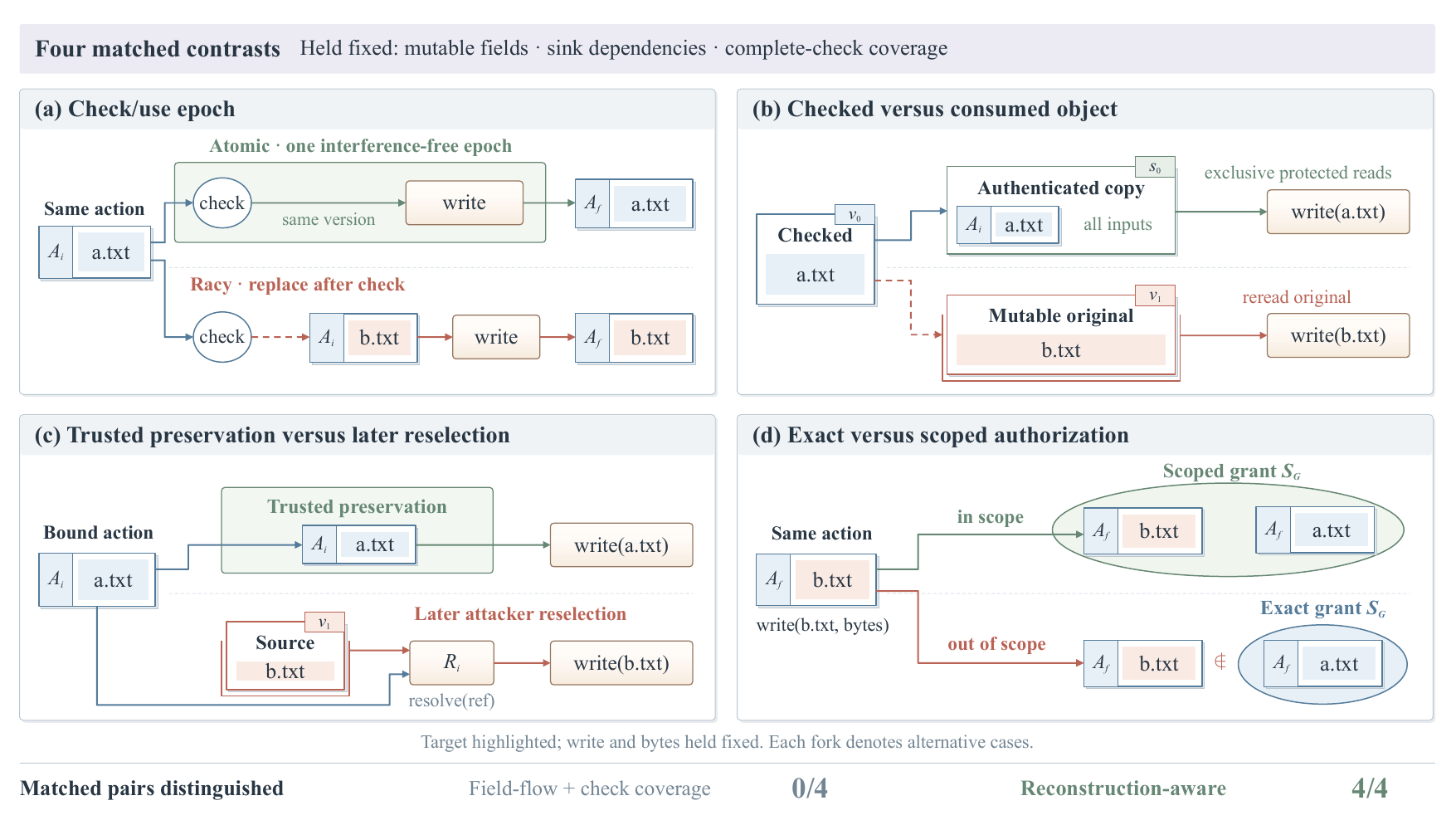}
\caption{Four authorization dimensions beyond field coverage: check/use epoch,
consumed object, later reconstruction, and grant-relative meaning. Changing
one dimension can reverse the repair judgment.}
\Description{The top timeline distinguishes the check/use epoch and the action
version consumed at use. Three lower comparisons show consumption of an
exclusive copy versus a mutable reread, trusted preservation versus later
reselection, and a changed target that remains within a scoped grant versus the
same change falling outside an exact grant.}
\label{fig:analysis-dimensions}
\end{figure}

For a competitive baseline, CodeQL 2.27.0 extracts dominance, object labels, trusted-copy calls, intervening writes, and candidate targets. Both analyzers receive the same fields, grant scope, and check/consume APIs; object relations, atomicity verdicts, labels, residuals, and outcomes are withheld. With the shared grant contract, the composite also scores \code{8/8}, separates \code{4/4}, and reports all unsafe residuals. This result locates the contribution precisely: conventional components can implement the judgment when APIs are explicit, while ReSA supplies the approval-consumer specification and repair diagnostic. Figure~\ref{fig:analysis-dimensions} shows the four isolated dimensions.

The unified \code{extract--validate--analyze} CLI uses these components as its front end. Entry-point reachability and authorization-context ranking reduce the displayed review inventory to 183 functions, 321 call edges, and 2,487 field accesses across the four packages; the full inventory remains available for recall auditing. On the released graph inputs, its shallow AST extractor recovers \code{27/35} transition edges and \code{45/233} edge--field facts; the CodeQL structural queries recover \code{26/35} edges and \code{104/233} facts. These extractors do not receive the grant contract and therefore do not emit repair verdicts. We report them as inputs to the analysis comparison, not as failed end-to-end competitors: scope, trusted preservation, atomicity, and opaque cross-representation links remain semantic facts that neither extractor infers reliably.

\subsubsection{Effect of obligation and projection omission}
We next perturb protections and graph inputs rather than add more topologies. Table~\ref{tab:rq2-robustness} reports the complete RQ2 matrix. Projection omission removes one graph-declared field from the final check; obligation removal disables one executor-side edge--field protection while retaining its analysis record. The hidden-field test adds a mutable, sink-influencing \code{policy_context} dependency to graph facts but omits it from the approval projection. Thus, its abstention tests fail-closed handling of a visible dependency, not discovery of an unknown schema field.

\subsubsection{Effect of graph-fact errors}
The graph-fact sensitivity test perturbs one declared fact at a time. Its three false-safe outcomes identify omitted reconstruction, mislabelled attacker control, and false trusted preservation as review-critical inputs; a visible projection omission instead produces a safe abstention. These tests measure the consequences of fact errors, not their prevalence.

\begin{table*}[t]
\centering
\caption{RQ2 abstraction, omission, and graph-input sensitivity results. A
case is one frozen graph or matched configuration; executions are independent
mutations or controls scored against final sink effects.}
\label{tab:rq2-robustness}
\scriptsize
\setlength{\tabcolsep}{3pt}
\begin{tabular}{p{.19\textwidth} p{.15\textwidth} p{.28\textwidth} p{.29\textwidth}}
\toprule
Experiment & Scale & Observed result & What the result tests \\
\midrule
Matched abstraction pairs & 4 pairs / 8 cases & Coverage \code{4/8}, \code{0/4} pairs; ours and CodeQL composite \code{8/8}, \code{4/4} pairs & Need for authorization facts beyond field coverage; conventional components realize them when APIs are explicit \\
Tool-assisted fact recovery & 4 released consumers & AST: edges \code{27/35}, facts \code{45/233}; CodeQL: edges \code{26/35}, facts \code{104/233} & How much of the analyst-provided graph can be recovered before semantic review \\
Projection omission & 17 cases / 85 mutations & Residual set exact \code{17/17}; omitted field reached \code{20/20}, included fields blocked \code{65/65} & Whether repair analysis identifies the field left outside an otherwise complete projection \\
Obligation removal & 20 interventions & Predicted candidate exposed \code{20/20}; exact controls valid \code{20/20} & Whether each modeled protection matters with all other protections retained \\
Visible hidden field & 4 graphs / 8 executions & Abstained \code{4/4}; ignoring abstention reached \code{4/4}; exact valid \code{4/4} & Fail-closed response when graph facts expose a sink dependency omitted from approval \\
Graph-fact perturbation & 6 inputs & 3 false-safe, 1 safe abstention, 2 correct & Sensitivity to omitted reconstruction, attacker control, trusted preservation, and unrelated atomicity \\
Independent graph construction & 2 analysts / 4 graphs & Dispositions \code{4/4}; sink dependencies \code{19/20} (F1 \code{0.974}); 1 valid + 3 abstentions each; median \code{63} min & Model-validation reproducibility and review cost \\
\bottomrule
\end{tabular}
\end{table*}

Both analysts accept C03 and abstain on C01, C02, and C04, yielding \code{4/4} model-validation disposition agreement. After canonical field names are aligned at the shared authority-bearing sinks, their sink-dependency sets overlap on \code{19} of the \code{20} fields in their union (precision \code{1.00}, recall \code{0.95}). Different graph boundaries leave only five matching edge--field keys across 95 and 79 facts, so a pooled field-label $\kappa$ would be uninformative. Median active review time is \code{63} minutes per case (range \code{37--72}).

\paragraph{Answer to RQ2.}
Field coverage alone cannot determine repair outcome: matched cases with the same fields and checks diverge when object version, check/use epoch, trusted preservation, later selection, or grant scope changes. Protection removal exposes the predicted path, while errors in three security-critical graph facts can produce false safety. A CodeQL composite recovers all eight source-level judgments when these APIs and the grant contract are explicit. Independent modeling reproduces all four model-validation dispositions and \code{19/20} sink dependencies.

\subsection{Transfer to Released Consumers (RQ3)}
RQ3 asks whether predictions frozen from released source paths survive actual execution. We evaluate sink effects, installed repair controls, and blind source/trace localization separately, so that each conclusion follows from the corresponding evidence class.

\subsubsection{Released paths and observed effects}
The direct/simulated-sink tier contains three witnesses. \pkg{md-feedback@1.5.6} carries \code{20/20} mutations to its package-owned file sink; PromptSpeak \code{0.3.0} carries \code{10/10} target/content mutations to its simulated executor and neutralizes the other \code{15/25}; Lobster \code{2026.9.7} carries \code{15/15} field mutations and \code{5/5} self-mutations through its published executor to a synthetic child-process sink. Exact and no-grant controls behave as expected.

The AI SDK path is configuration-sensitive: without its documented secret, \code{15/15} mutations reach the Octokit recorder; with HMAC, exact execution succeeds in \code{5/5} and all mutations fail before executor entry. Signed replay remains possible in \code{5/5}, separating integrity from freshness. Agent Approve and Ng-annotate provide delegated effects; Pincushion is a packet-only negative whose final gate blocks impact.

OpenClaw is the equality hard negative: \code{10/10} target/content changes remain within directory scope, \code{5/5} operation changes are rejected, and APAS is \code{0/15}. Its independent scope checker passes \code{69/69} cases. Eve and Patchwork retain the approved action and neutralize changed responses.

Three releases broaden the controls. With \pkg{@sema-agent/core@7.10.0}, its detached snapshot preserves the reviewed write in \code{5/5}, whereas consuming a mutable alias carries the attacker write in \code{5/5}. With \pkg{ssh-mcp@2.8.0}, the released transport rejects changed and expired commands; against an isolated SSH service, rereading mutable input creates the attacker marker in \code{5/5}, while consuming the checked command creates only the reviewed marker. With \pkg{@voyant-travel/action-ledger@0.115.9}, the complete fingerprint distinguishes \code{15/15} mutations, whereas a coarse projection collides on \code{10/10} omitted-component mutations. Predictions match all five structural cases.

\subsubsection{Installed repairs and blind transfer}
Table~\ref{tab:repair} reports five repair controls and a PromptSpeak insufficient-repair contrast. Sufficient interventions preserve exact execution and prevent \code{60/60} action mutations: 55 rejected and five snapshot-neutralized. The denominator is \code{15+20+5+5+15}; caller controls and signed replay are reported separately. The prospective configurations provide the cross-project test.

\begin{table*}[t]
\centering
\caption{Released-consumer repair evidence. ``Installed'' means the reference
intervention was executed in the experiment; blind localization is reported
separately in text.}
\label{tab:repair}
\scriptsize
\setlength{\tabcolsep}{3pt}
\begin{tabular}{p{.16\textwidth} p{.17\textwidth} p{.17\textwidth} p{.12\textwidth} p{.12\textwidth} p{.17\textwidth}}
\toprule
Consumer & Sink evidence & Predicted intervention & Installed & Exact & Invalid-action outcome \\
\midrule
\pkg{md-feedback@1.5.6} & Package-owned file sink & Atomic canonical digest before grant consumption & yes, isolated fork & \code{5/5} & rejected \code{15/15} \\
PromptSpeak \code{0.3.0} & Package-owned simulated executor & Complete arguments and caller binding & yes, isolated fork & \code{5/5} & rejected \code{20/20} \\
Lobster \code{2026.9.7} & Released executor; synthetic child sink & Consume attacker-immutable checked snapshot & yes, wrapper & \code{5/5} & reached \code{5/5} before; neutralized \code{5/5} after \\
Agent Approve \code{0.1.24} & Delegated OpenCode hook & Atomic tool-and-argument digest & yes, wrapper & \code{5/5} & rejected \code{5/5} \\
AI SDK \code{7.0.58} & Released SDK; local recorder & Documented HMAC over tool and input & yes, SDK path & \code{5/5} & rejected \code{15/15}; replay separate \\
PromptSpeak partial binding & Same simulated executor & Target-only binding (predicted insufficient) & yes, isolated fork & \code{5/5} & content reached \code{5/5}; complete binding rejected \code{5/5} \\
\midrule
Transfer total & Five repair controls & Frozen sufficient interventions & \code{5/5} & \code{5/5} & prevented \code{60/60}: 55 rejected, five neutralized \\
\bottomrule
\end{tabular}
\end{table*}

The blind set validates topology, classification, boundary, and source/trace repair localization on \code{3/3} version-pinned consumers; it does not install repairs. A clean rerun reproduces all counts, including \code{12/12} vulnerable and \code{6/6} neutralized blind rows and Lobster's \code{5/5} before versus \code{0/5} after snapshot consumption.

The released-code evidence is purposive rather than representative. We distinguish three blind consumers, five executed repair controls, and three residual-effect contrasts; packages without a complete approval-to-sink path are not treated as safe negatives. The artifact records version pins, inclusion criteria, source anchors, and exclusion reasons.

\paragraph{Answer to RQ3.}
Frozen diagnoses transfer to all three blind consumers. Five executed controls realize the predicted safe outcome. Three mechanism contrasts leave the predicted residual effects reachable: PromptSpeak partial binding, Lobster check-then-reread, and SSH check-then-reread.

\section{Related Work}
Loopjacking studies approval-to-effect mismatch in released agent products, including incomplete approval representations and post-approval state substitution~\cite{loopjacking}. APAS-Finder complements this product-level perspective with repair-oriented reasoning over existing consumers. Given representation transitions, sink dependencies, consumed object versions, and authorization scope, it derives obligations, evaluates candidate repairs, and identifies the residual sink suffix when a repair does not cover the reconstructed action.

Complete mediation, least privilege, access-control calculi, and confused-deputy work establish when authority must be checked and how it is delegated~\cite{saltzerschroeder,abadicalculus,confuseddeputy}. Recent agent-security studies demonstrate execution-boundary failures and indirect prompt injection in tool-integrated systems~\cite{agentbound,greshake,agentdojo,injecagent}. The closest concurrent proposals, Consent Integrity and CONTINUITY, bind approved actions or security context across execution transitions~\cite{consentintegrity,continuity}. ReSA asks a complementary implementation-facing question: whether an existing consumer preserves that binding while reconstructing the action and, if not, which repair leaves a residual sink suffix.

Program analyses provide the building blocks: interprocedural reachability, object-sensitive points-to analysis, information flow, and typestate recover dependencies, identities, and protocol states~\cite{repsifds,milanovaobjectsensitivity,sabelfeldmyers,stromtypestate,delinetypestate,livshitsstaticsecurity}. ReSA combines reviewed facts into obligations, verifies the consumed version, and reports repair-specific suffixes. This composition separates the matched cases in Figure~\ref{fig:analysis-dimensions} despite identical field-flow and check coverage. Relative to CONTINUITY's linter, it outputs the object/version condition required by a repair and the residual path left in an existing consumer.

A backward slice recovers candidate data paths, while object-sensitive points-to, dominance, lockset, and typestate analyses populate model facts. ReSA relates those paths to approved scope, semantic preservation, and grant-equivalent objects. Its suffix is conditioned on grant $G$ and repair $H$ and begins at an undischarged obligation. The CodeQL evaluation separately measures graph-fact recovery and an explicit-API composite over the eight matched cases.

Runtime monitors and edit automata establish how event policies can mediate execution~\cite{schneiderpolicies,ligattiedit}. ReSA specializes that enforcement question to approval consumers: it locates the consumed object, determines whether preservation or atomic validation discharges each obligation, and reports the remaining suffix.

Linearizability and file-race analyses supply the established basis for the check/use condition~\cite{herlihywing,lheechapin,weiputocttou}. ReSA combines that condition with approval scope and object custody to judge a candidate repair.

AgentDojo, InjecAgent, and ToolSafe exercise prompt injection, changing state, and guarded tool invocation~\cite{agentdojo,injecagent,toolsafe}. ReSA instead derives repair obligations tied to the consumed action before execution. Following repair-correctness work~\cite{monperrusrepair,smithoverfitting,xiongpatchcorrectness}, the evaluation checks both rejection of out-of-scope effects and preservation of exact approved behavior.

Replay and duplicate execution are freshness concerns separate from the relation studied here: whether the action selected after approval remains inside the approved projection.

\section{Discussion and Limitations}
Three boundaries shape the interpretation. First, dynamic evidence establishes reachability rather than human deception: deterministic mutations define the oracle, and approvals are synthetic. Sink evidence also ranges from package-owned effects to binding-only observations. Second, the analysis is relative to a reviewed graph. AST and CodeQL recover 27/35 and 26/35 edges, but scope, trust, atomicity, opaque flow, and omissions require review. The analysts agree on all four model-validation dispositions and \code{19/20} sink dependencies, yet choose boundaries with too few common edge--field keys for a meaningful pooled $\kappa$. Four cases and self-reported timing do not establish population-level annotation reliability or unattended use.

Third, the semantics cover one exact or scoped grant rather than multi-action plans. The study characterizes the tested repairs and consumers rather than ecosystem prevalence, omitted fields, policy errors, or key compromise.

\paragraph{Ethics.} The modeling study retains anonymized technical annotations and self-reported task time from two analysts, but no demographic or sensitive personal data. All software effects remain in isolated local targets or recorders, and repairs are not maintainer-confirmed.

\section{Conclusion}
We address approval-to-execution drift along three dimensions: ReSA defines the required semantic relation, APAS-Finder derives repair obligations and residual suffixes, and independent sink execution tests those predictions. Predictions match all 28 controlled cases; five controls prevent 60 tested out-of-scope effects, while three mechanism contrasts retain the predicted residual effects. Across the selected released paths, the analysis identifies what each repair must protect and where an uncovered path remains.

The practical message is simple: bind the action that will be consumed, not merely the approval record that preceded it.

\section*{Data Availability}
An anonymous package provides frozen annotator packets and submissions, source anchors, validation and agreement scripts, executors, observations, scoring, and integrity metadata; finite-model claims remain declaration-relative.

\bibliographystyle{ACM-Reference-Format}
\bibliography{main}


\begin{thebibliography}{27}


\ifx \showCODEN    \undefined \def \showCODEN     #1{\unskip}     \fi
\ifx \showISBNx    \undefined \def \showISBNx     #1{\unskip}     \fi
\ifx \showISBNxiii \undefined \def \showISBNxiii  #1{\unskip}     \fi
\ifx \showISSN     \undefined \def \showISSN      #1{\unskip}     \fi
\ifx \showLCCN     \undefined \def \showLCCN      #1{\unskip}     \fi
\ifx \shownote     \undefined \def \shownote      #1{#1}          \fi
\ifx \showarticletitle \undefined \def \showarticletitle #1{#1}   \fi
\ifx \showURL      \undefined \def \showURL       {\relax}        \fi
\providecommand\bibfield[2]{#2}
\providecommand\bibinfo[2]{#2}
\providecommand\natexlab[1]{#1}
\providecommand\showeprint[2][]{arXiv:#2}
\makeatletter
\@ifundefined{NAT@parse@date}{}{\let\NAT@parse@date@orig\NAT@parse@date}
\@ifundefined{NAT@parse@date}{}{\def\NAT@parse@date#1#2#3#4#5#6@@{\NAT@parse@date@orig#1#2#3#4#5#6@@\def\NAT@tempyear{0000}\def\NAT@tempexlab{{?}}\ifx\NAT@year\NAT@tempyear\ifx\NAT@exlab\NAT@tempexlab\def\NAT@date{[n.\,d.]}\else\edef\NAT@date{[n.\,d.]\NAT@exlab}\fi\fi}}
\makeatother

\bibitem[Abadi et~al\mbox{.}(1993)]%
        {abadicalculus}
\bibfield{author}{\bibinfo{person}{Mart\'in Abadi}, \bibinfo{person}{Michael
  Burrows}, \bibinfo{person}{Butler Lampson}, {and} \bibinfo{person}{Gordon
  Plotkin}.} \bibinfo{year}{1993}\natexlab{}.
\newblock \showarticletitle{A Calculus for Access Control in Distributed
  Systems}.
\newblock \bibinfo{journal}{\emph{ACM Transactions on Programming Languages and
  Systems}} \bibinfo{volume}{15}, \bibinfo{number}{4} (\bibinfo{year}{1993}),
  \bibinfo{pages}{706--734}.
\newblock
\href{https://doi.org/10.1145/155183.155225}{doi:\nolinkurl{10.1145/155183.155225}}


\bibitem[B\"uhler et~al\mbox{.}(2026)]%
        {agentbound}
\bibfield{author}{\bibinfo{person}{Christoph B\"uhler}, \bibinfo{person}{Matteo
  Biagiola}, \bibinfo{person}{Luca Di~Grazia}, {and} \bibinfo{person}{Guido
  Salvaneschi}.} \bibinfo{year}{2026}\natexlab{}.
\newblock \showarticletitle{AgentBound: Securing Execution Boundaries of AI
  Agents}.
\newblock \bibinfo{journal}{\emph{Proceedings of the ACM on Software
  Engineering}} (\bibinfo{year}{2026}).
\newblock
\href{https://doi.org/10.1145/3808103}{doi:\nolinkurl{10.1145/3808103}}


\bibitem[Debenedetti et~al\mbox{.}(2024)]%
        {agentdojo}
\bibfield{author}{\bibinfo{person}{Edoardo Debenedetti}, \bibinfo{person}{Jie
  Zhang}, \bibinfo{person}{Mislav Balunovic}, \bibinfo{person}{Luca
  Beurer-Kellner}, \bibinfo{person}{Marc Fischer}, {and}
  \bibinfo{person}{Florian Tram\`er}.} \bibinfo{year}{2024}\natexlab{}.
\newblock \showarticletitle{AgentDojo: A Dynamic Environment to Evaluate Prompt
  Injection Attacks and Defenses for LLM Agents}. In
  \bibinfo{booktitle}{\emph{Advances in Neural Information Processing Systems
  37}}. \bibinfo{pages}{82895--82920}.
\newblock
\href{https://doi.org/10.52202/079017-2636}{doi:\nolinkurl{10.52202/079017-2636}}


\bibitem[DeLine and F\"ahndrich(2004)]%
        {delinetypestate}
\bibfield{author}{\bibinfo{person}{Robert DeLine} {and} \bibinfo{person}{Manuel
  F\"ahndrich}.} \bibinfo{year}{2004}\natexlab{}.
\newblock \showarticletitle{Typestates for Objects}. In
  \bibinfo{booktitle}{\emph{Proceedings of the 18th European Conference on
  Object-Oriented Programming}}. \bibinfo{pages}{465--490}.
\newblock
\href{https://doi.org/10.1007/978-3-540-24851-4_21}{doi:\nolinkurl{10.1007/978-3-540-24851-4_21}}


\bibitem[Fisler et~al\mbox{.}(2005)]%
        {fisleraccesscontrol}
\bibfield{author}{\bibinfo{person}{Kathi Fisler}, \bibinfo{person}{Shriram
  Krishnamurthi}, \bibinfo{person}{Leo~A. Meyerovich}, {and}
  \bibinfo{person}{Michael~Carl Tschantz}.} \bibinfo{year}{2005}\natexlab{}.
\newblock \showarticletitle{Verification and Change-Impact Analysis of
  Access-Control Policies}. In \bibinfo{booktitle}{\emph{Proceedings of the
  27th International Conference on Software Engineering}}.
  \bibinfo{pages}{196--205}.
\newblock
\href{https://doi.org/10.1145/1062455.1062502}{doi:\nolinkurl{10.1145/1062455.1062502}}


\bibitem[Greshake et~al\mbox{.}(2023)]%
        {greshake}
\bibfield{author}{\bibinfo{person}{Kai Greshake}, \bibinfo{person}{Saeed
  Abdelnabi}, \bibinfo{person}{Shailesh Mishra}, \bibinfo{person}{Christoph
  Endres}, \bibinfo{person}{Thorsten Holz}, {and} \bibinfo{person}{Mario
  Fritz}.} \bibinfo{year}{2023}\natexlab{}.
\newblock \showarticletitle{Not What You've Signed Up For: Compromising
  Real-World LLM-Integrated Applications with Indirect Prompt Injection}. In
  \bibinfo{booktitle}{\emph{Proceedings of the 16th ACM Workshop on Artificial
  Intelligence and Security}}.
\newblock
\href{https://doi.org/10.1145/3605764.3623985}{doi:\nolinkurl{10.1145/3605764.3623985}}


\bibitem[Hardy(1988)]%
        {confuseddeputy}
\bibfield{author}{\bibinfo{person}{Norm Hardy}.}
  \bibinfo{year}{1988}\natexlab{}.
\newblock \showarticletitle{The Confused Deputy: (or Why Capabilities Might
  Have Been Invented)}.
\newblock \bibinfo{journal}{\emph{Operating Systems Review}}
  \bibinfo{volume}{22}, \bibinfo{number}{4} (\bibinfo{year}{1988}),
  \bibinfo{pages}{36--38}.
\newblock
\href{https://doi.org/10.1145/54289.871709}{doi:\nolinkurl{10.1145/54289.871709}}


\bibitem[Herlihy and Wing(1990)]%
        {herlihywing}
\bibfield{author}{\bibinfo{person}{Maurice~P. Herlihy} {and}
  \bibinfo{person}{Jeannette~M. Wing}.} \bibinfo{year}{1990}\natexlab{}.
\newblock \showarticletitle{Linearizability: A Correctness Condition for
  Concurrent Objects}.
\newblock \bibinfo{journal}{\emph{ACM Transactions on Programming Languages and
  Systems}} \bibinfo{volume}{12}, \bibinfo{number}{3} (\bibinfo{year}{1990}),
  \bibinfo{pages}{463--492}.
\newblock
\href{https://doi.org/10.1145/78969.78972}{doi:\nolinkurl{10.1145/78969.78972}}


\bibitem[Jia and Harman(2011)]%
        {jiaharmanmutation}
\bibfield{author}{\bibinfo{person}{Yue Jia} {and} \bibinfo{person}{Mark
  Harman}.} \bibinfo{year}{2011}\natexlab{}.
\newblock \showarticletitle{An Analysis and Survey of the Development of
  Mutation Testing}.
\newblock \bibinfo{journal}{\emph{IEEE Transactions on Software Engineering}}
  \bibinfo{volume}{37}, \bibinfo{number}{5} (\bibinfo{year}{2011}),
  \bibinfo{pages}{649--678}.
\newblock
\href{https://doi.org/10.1109/TSE.2010.62}{doi:\nolinkurl{10.1109/TSE.2010.62}}


\bibitem[Lhee and Chapin(2005)]%
        {lheechapin}
\bibfield{author}{\bibinfo{person}{Kyung-suk Lhee} {and}
  \bibinfo{person}{Steve~J. Chapin}.} \bibinfo{year}{2005}\natexlab{}.
\newblock \showarticletitle{Detection of File-Based Race Conditions}.
\newblock \bibinfo{journal}{\emph{International Journal of Information
  Security}} \bibinfo{volume}{4}, \bibinfo{number}{1--2}
  (\bibinfo{year}{2005}), \bibinfo{pages}{105--119}.
\newblock
\href{https://doi.org/10.1007/s10207-004-0068-2}{doi:\nolinkurl{10.1007/s10207-004-0068-2}}


\bibitem[Ligatti et~al\mbox{.}(2005)]%
        {ligattiedit}
\bibfield{author}{\bibinfo{person}{Jay Ligatti}, \bibinfo{person}{Lujo Bauer},
  {and} \bibinfo{person}{David Walker}.} \bibinfo{year}{2005}\natexlab{}.
\newblock \showarticletitle{Edit Automata: Enforcement Mechanisms for Run-Time
  Security Policies}.
\newblock \bibinfo{journal}{\emph{International Journal of Information
  Security}} \bibinfo{volume}{4}, \bibinfo{number}{1--2}
  (\bibinfo{year}{2005}), \bibinfo{pages}{2--16}.
\newblock
\href{https://doi.org/10.1007/s10207-004-0046-8}{doi:\nolinkurl{10.1007/s10207-004-0046-8}}


\bibitem[Livshits and Lam(2005)]%
        {livshitsstaticsecurity}
\bibfield{author}{\bibinfo{person}{V.~Benjamin Livshits} {and}
  \bibinfo{person}{Monica~S. Lam}.} \bibinfo{year}{2005}\natexlab{}.
\newblock \showarticletitle{Finding Security Vulnerabilities in Java
  Applications with Static Analysis}. In \bibinfo{booktitle}{\emph{Proceedings
  of the 14th USENIX Security Symposium}}.
\newblock
\urldef\tempurl%
\url{https://www.usenix.org/conference/14th-usenix-security-symposium/finding-security-vulnerabilities-java-applications-static}
\showURL{%
\tempurl}


\bibitem[Milanova et~al\mbox{.}(2005)]%
        {milanovaobjectsensitivity}
\bibfield{author}{\bibinfo{person}{Ana Milanova}, \bibinfo{person}{Atanas
  Rountev}, {and} \bibinfo{person}{Barbara~G. Ryder}.}
  \bibinfo{year}{2005}\natexlab{}.
\newblock \showarticletitle{Parameterized Object Sensitivity for Points-to
  Analysis for Java}.
\newblock \bibinfo{journal}{\emph{ACM Transactions on Software Engineering and
  Methodology}} \bibinfo{volume}{14}, \bibinfo{number}{1}
  (\bibinfo{year}{2005}), \bibinfo{pages}{1--41}.
\newblock
\href{https://doi.org/10.1145/1044834.1044835}{doi:\nolinkurl{10.1145/1044834.1044835}}


\bibitem[Monperrus(2018)]%
        {monperrusrepair}
\bibfield{author}{\bibinfo{person}{Martin Monperrus}.}
  \bibinfo{year}{2018}\natexlab{}.
\newblock \showarticletitle{Automatic Software Repair: A Bibliography}.
\newblock \bibinfo{journal}{\emph{Comput. Surveys}} \bibinfo{volume}{51},
  \bibinfo{number}{1} (\bibinfo{year}{2018}), \bibinfo{pages}{1--24}.
\newblock
\href{https://doi.org/10.1145/3105906}{doi:\nolinkurl{10.1145/3105906}}


\bibitem[Mou et~al\mbox{.}(2026)]%
        {toolsafe}
\bibfield{author}{\bibinfo{person}{Yutao Mou}, \bibinfo{person}{Zhangchi Xue},
  \bibinfo{person}{Lijun Li}, \bibinfo{person}{Peiyang Liu},
  \bibinfo{person}{Shikun Zhang}, \bibinfo{person}{Wei Ye}, {and}
  \bibinfo{person}{Jing Shao}.} \bibinfo{year}{2026}\natexlab{}.
\newblock \showarticletitle{ToolSafe: Enhancing Tool Invocation Safety of
  LLM-based Agents via Proactive Step-level Guardrail and Feedback}. In
  \bibinfo{booktitle}{\emph{Findings of the Association for Computational
  Linguistics: ACL 2026}}.
\newblock
\href{https://doi.org/10.18653/v1/2026.findings-acl.1850}{doi:\nolinkurl{10.18653/v1/2026.findings-acl.1850}}


\bibitem[Reps et~al\mbox{.}(1995)]%
        {repsifds}
\bibfield{author}{\bibinfo{person}{Thomas Reps}, \bibinfo{person}{Susan
  Horwitz}, {and} \bibinfo{person}{Mooly Sagiv}.}
  \bibinfo{year}{1995}\natexlab{}.
\newblock \showarticletitle{Precise Interprocedural Dataflow Analysis via Graph
  Reachability}. In \bibinfo{booktitle}{\emph{Proceedings of the 22nd ACM
  SIGPLAN-SIGACT Symposium on Principles of Programming Languages}}.
  \bibinfo{pages}{49--61}.
\newblock
\href{https://doi.org/10.1145/199448.199462}{doi:\nolinkurl{10.1145/199448.199462}}


\bibitem[Sabelfeld and Myers(2003)]%
        {sabelfeldmyers}
\bibfield{author}{\bibinfo{person}{Andrei Sabelfeld} {and}
  \bibinfo{person}{Andrew~C. Myers}.} \bibinfo{year}{2003}\natexlab{}.
\newblock \showarticletitle{Language-Based Information-Flow Security}.
\newblock \bibinfo{journal}{\emph{IEEE Journal on Selected Areas in
  Communications}} \bibinfo{volume}{21}, \bibinfo{number}{1}
  (\bibinfo{year}{2003}), \bibinfo{pages}{5--19}.
\newblock
\href{https://doi.org/10.1109/JSAC.2002.806121}{doi:\nolinkurl{10.1109/JSAC.2002.806121}}


\bibitem[Saltzer and Schroeder(1975)]%
        {saltzerschroeder}
\bibfield{author}{\bibinfo{person}{Jerome~H. Saltzer} {and}
  \bibinfo{person}{Michael~D. Schroeder}.} \bibinfo{year}{1975}\natexlab{}.
\newblock \showarticletitle{The Protection of Information in Computer Systems}.
\newblock \bibinfo{journal}{\emph{Proc. IEEE}} \bibinfo{volume}{63},
  \bibinfo{number}{9} (\bibinfo{year}{1975}), \bibinfo{pages}{1278--1308}.
\newblock
\href{https://doi.org/10.1109/PROC.1975.9939}{doi:\nolinkurl{10.1109/PROC.1975.9939}}


\bibitem[Schneider(2000)]%
        {schneiderpolicies}
\bibfield{author}{\bibinfo{person}{Fred~B. Schneider}.}
  \bibinfo{year}{2000}\natexlab{}.
\newblock \showarticletitle{Enforceable Security Policies}.
\newblock \bibinfo{journal}{\emph{ACM Transactions on Information and System
  Security}} \bibinfo{volume}{3}, \bibinfo{number}{1} (\bibinfo{year}{2000}),
  \bibinfo{pages}{30--50}.
\newblock
\href{https://doi.org/10.1145/353323.353382}{doi:\nolinkurl{10.1145/353323.353382}}


\bibitem[Schwartz et~al\mbox{.}(2010)]%
        {schwartztaint}
\bibfield{author}{\bibinfo{person}{Edward~J. Schwartz},
  \bibinfo{person}{Thanassis Avgerinos}, {and} \bibinfo{person}{David
  Brumley}.} \bibinfo{year}{2010}\natexlab{}.
\newblock \showarticletitle{All You Ever Wanted to Know about Dynamic Taint
  Analysis and Forward Symbolic Execution (but Might Have Been Afraid to Ask)}.
  In \bibinfo{booktitle}{\emph{2010 IEEE Symposium on Security and Privacy}}.
  \bibinfo{pages}{317--331}.
\newblock
\href{https://doi.org/10.1109/SP.2010.26}{doi:\nolinkurl{10.1109/SP.2010.26}}


\bibitem[Smith et~al\mbox{.}(2015)]%
        {smithoverfitting}
\bibfield{author}{\bibinfo{person}{Edward~K. Smith}, \bibinfo{person}{Earl~T.
  Barr}, \bibinfo{person}{Claire Le~Goues}, {and} \bibinfo{person}{Yuriy
  Brun}.} \bibinfo{year}{2015}\natexlab{}.
\newblock \showarticletitle{Is the Cure Worse than the Disease? Overfitting in
  Automated Program Repair}. In \bibinfo{booktitle}{\emph{Proceedings of the
  10th Joint Meeting on Foundations of Software Engineering}}.
  \bibinfo{pages}{532--543}.
\newblock
\href{https://doi.org/10.1145/2786805.2786825}{doi:\nolinkurl{10.1145/2786805.2786825}}


\bibitem[Strom and Yemini(1986)]%
        {stromtypestate}
\bibfield{author}{\bibinfo{person}{Robert~E. Strom} {and}
  \bibinfo{person}{Shaula Yemini}.} \bibinfo{year}{1986}\natexlab{}.
\newblock \showarticletitle{Typestate: A Programming Language Concept for
  Enhancing Software Reliability}.
\newblock \bibinfo{journal}{\emph{IEEE Transactions on Software Engineering}}
  \bibinfo{volume}{SE-12}, \bibinfo{number}{1} (\bibinfo{year}{1986}),
  \bibinfo{pages}{157--171}.
\newblock
\href{https://doi.org/10.1109/TSE.1986.6312929}{doi:\nolinkurl{10.1109/TSE.1986.6312929}}


\bibitem[Wei and Pu(2010)]%
        {weiputocttou}
\bibfield{author}{\bibinfo{person}{Jinpeng Wei} {and} \bibinfo{person}{Calton
  Pu}.} \bibinfo{year}{2010}\natexlab{}.
\newblock \showarticletitle{Modeling and Preventing {TOCTTOU} Vulnerabilities
  in {Unix}-Style File Systems}.
\newblock \bibinfo{journal}{\emph{Computers \& Security}} \bibinfo{volume}{29},
  \bibinfo{number}{8} (\bibinfo{year}{2010}), \bibinfo{pages}{815--830}.
\newblock
\href{https://doi.org/10.1016/j.cose.2010.09.004}{doi:\nolinkurl{10.1016/j.cose.2010.09.004}}


\bibitem[Weng(2026)]%
        {consentintegrity}
\bibfield{author}{\bibinfo{person}{Xiaoqi Weng}.}
  \bibinfo{year}{2026}\natexlab{}.
\newblock \bibinfo{title}{What You Approve Is What Executes: Consent Integrity
  for Black-Box LLM Agents}.
\newblock
\showeprint[arxiv]{2606.02668}


\bibitem[Xiong et~al\mbox{.}(2018)]%
        {xiongpatchcorrectness}
\bibfield{author}{\bibinfo{person}{Yingfei Xiong}, \bibinfo{person}{Xinyuan
  Liu}, \bibinfo{person}{Muhan Zeng}, \bibinfo{person}{Lu Zhang}, {and}
  \bibinfo{person}{Gang Huang}.} \bibinfo{year}{2018}\natexlab{}.
\newblock \showarticletitle{Identifying Patch Correctness in Test-Based Program
  Repair}. In \bibinfo{booktitle}{\emph{Proceedings of the 40th International
  Conference on Software Engineering}}. \bibinfo{pages}{789--799}.
\newblock
\href{https://doi.org/10.1145/3180155.3180182}{doi:\nolinkurl{10.1145/3180155.3180182}}


\bibitem[Zhan et~al\mbox{.}(2024)]%
        {injecagent}
\bibfield{author}{\bibinfo{person}{Qiusi Zhan}, \bibinfo{person}{Zhixiang
  Liang}, \bibinfo{person}{Zifan Ying}, {and} \bibinfo{person}{Daniel Kang}.}
  \bibinfo{year}{2024}\natexlab{}.
\newblock \showarticletitle{InjecAgent: Benchmarking Indirect Prompt Injections
  in Tool-Integrated Large Language Model Agents}. In
  \bibinfo{booktitle}{\emph{Findings of the Association for Computational
  Linguistics: ACL 2024}}. \bibinfo{pages}{10471--10506}.
\newblock
\href{https://doi.org/10.18653/v1/2024.findings-acl.624}{doi:\nolinkurl{10.18653/v1/2024.findings-acl.624}}


\bibitem[Zheng and Yang(2026)]%
        {continuity}
\bibfield{author}{\bibinfo{person}{Chris Zheng} {and} \bibinfo{person}{Geng
  Yang}.} \bibinfo{year}{2026}\natexlab{}.
\newblock \bibinfo{title}{CONTINUITY: Security-Context Contracts for Composable
  LLM Agent Controls}.
\newblock
\showeprint[arxiv]{2609.05269}

\bibitem[Arun Kumar(2026)]%
        {loopjacking}
\bibfield{author}{\bibinfo{person}{Adithyan Arun Kumar}.}
  \bibinfo{year}{2026}\natexlab{}.
\newblock \bibinfo{title}{Loopjacking: Hijacking Human-in-the-Loop Approval}.
\newblock
\showeprint[arxiv]{2609.21081}



\end{thebibliography}
\end{document}